\documentclass[sigconf,manuscript]{acmart}
\AtBeginDocument{%
  \providecommand\BibTeX{{%
    \normalfont B\kern-0.5em{\scshape i\kern-0.25em b}\kern-0.8em\TeX}}}

\setcopyright{iw3c2w3g}
\copyrightyear{2021}
\acmDOI{}

\acmConference[KidRec'21]{KidRec '21: 5th International and Interdisciplinary Perspectives on Children \& Recommender and Information Retrieval Systems (KidRec) Search and Recommendation Technology through the Lens of a Teacher - Co-located with ACM IDC 2021}{June 26, 2021}{Online Event}

\usepackage{subfigure}

\begin{document}

\title[Pink for Princesses, Blue for Superheroes]{Pink for Princesses, Blue for Superheroes: The Need to Examine Gender Stereotypes in Kids' Products in Search and Recommendations}

\author{Amifa Raj}
\email{amifaraj@u.boisestate.edu}
\affiliation{
  \institution{People and Information Research Team, Boise State University}
  \city{Boise}
  \state{Idaho}
  \country{USA}
}
\author{Ashlee Milton}
\email{AshleeMilton@u.boisestate.edu}
\affiliation{
  \institution{People and Information Research Team, Boise State University}
  \city{Boise}
  \state{Idaho}
  \country{USA}
}

\author{Michael D. Ekstrand}
\email{michaelekstrand@boisestate.edu}
\affiliation{
  \institution{People and Information Research Team, Boise State University}
  \city{Boise}
  \state{Idaho}
  \country{USA}
}

\renewcommand{\shortauthors}{A. Raj et al.}

\begin{abstract}
In this position paper, we argue for the need to investigate if and how gender stereotypes manifest in search and recommender systems. As a starting point, we particularly focus on how these systems may propagate and reinforce gender stereotypes through their results in learning environments, a context where teachers and children in their formative stage regularly interact with these systems. We provide motivating examples supporting our concerns and outline an agenda to support future research addressing the phenomena.



\end{abstract}





\maketitle

\section{Introduction}

Gender stereotypes are common beliefs and social expectations associated with specific genders \cite{ellemers2018gender}. For example, \textit{blue} is often associated with boys' toys, whereas \textit{pink} is used for girls' toys \cite{fulcher2018building}. Including unnecessary stereotypes, particularly in learning environments, can affect children's ideas and future beliefs about gender roles \cite{seitz2020effects}. Empirical studies show that early exposure of gender stereotypes can not only have negative effects on children's ideas of gender roles, but also their creativity, confidence, skills, and accomplishments \cite{fulcher2018building, kneeskern2020examining}. While the impact of gender stereotypes on children has been heavily studied in psychology \cite{dinella2018gender}, there has yet to be significant work on the role search and recommendations may play in propagating these stereotypes to children in learning environments. Further, the extent to which search and recommender systems help or hinder teachers who work to create stereotype-free classroom is still unexplored. 
 
Both teachers and students often use search engines and recommender systems in classroom settings to facilitate early education \cite{azpiazu2017online}. As a consequence, if these systems reproduce gender stereotypes in such contexts, they may play a role in early childhood exposure to and reinforcement of gender stereotypes. This can be either through children's direct exposure as they use the systems themselves, or indirectly as teachers use the systems to locate learning resources for their students. \citet{gray2004perpetuating} and \citet{lo2015teachers} analyze school teacher perspectives towards gender stereotypes in classroom settings showing that teachers are generally aware of stereotypes as a general issue, but there has not yet been research on how teachers who practice gender equality in their classroom understand and view their encounters with gender stereotypes through search and recommendation systems specifically.
The role of search and recommendations in propagating other harmful expectations and beliefs to children is currently overlooked; filling this gap will help provide children with effective and equitable learning environments enhanced by search and recommendations.

In this paper, we argue that the research community needs to consider and address the possibility of search and recommendations systems propagating gender stereotypes to children in the classroom. That is, researchers and developers need to investigate whether and how these systems may perpetuate or reinforce gender stereotypes, develop strategies for detecting and measuring such stereotypes, and --- when present --- mitigate the gender stereotypes presented to children through search results and recommendations. This attention to gender stereotypes will complement existing work on improving both learning \citep{zeniarja2018search} and teaching \citep{ekstrand2020enhancing} with search and recommendations. We support our position with real-life examples showing that search and recommendation systems can replicate gender stereotypes associated with children’s products, exposing them to children. We outline several potential directions for further research through proposed research questions. We hope to spur work presenting methods to measure these stereotypes and identify causes, impacts, and mitigation strategies for early childhood exposure of gender stereotypes through search and recommendations.

\section{Examples of Gender Stereotypical Responses}

\begin{figure}[!ht]
    \subfigure[\label{fig:Amazon-Kid-result}Top-result]{
        \includegraphics[width=0.2\linewidth]{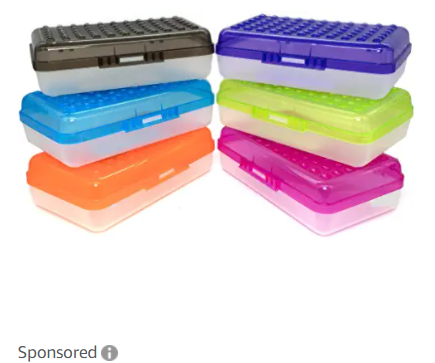}
    }
    \subfigure[\label{fig:Amazon-kid-Rec}Associated recommendations]{
        \includegraphics[width=0.4\linewidth]{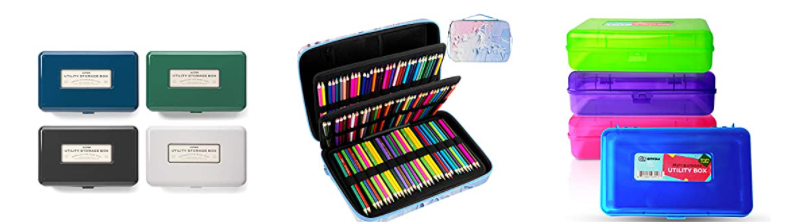}
    }
    \subfigure[\label{fig:girl}``Pencil box for girls"]{
        \includegraphics[width=0.4\linewidth]{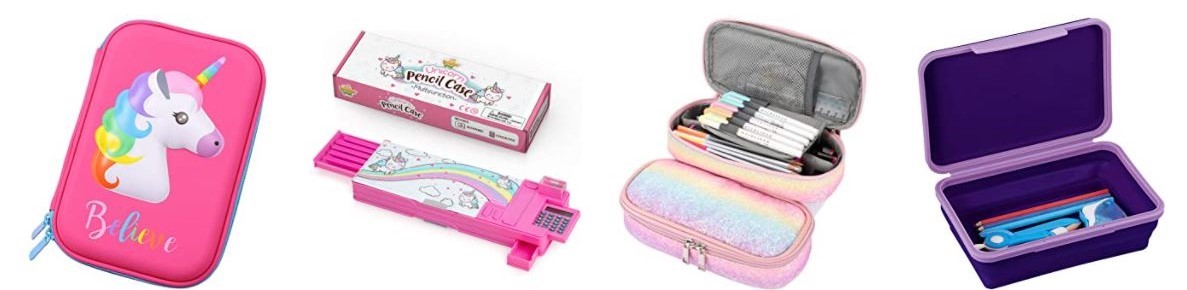}
    }
    \subfigure[\label{fig:boy}``Pencil box for boys"]{
        \includegraphics[width=0.4\linewidth]{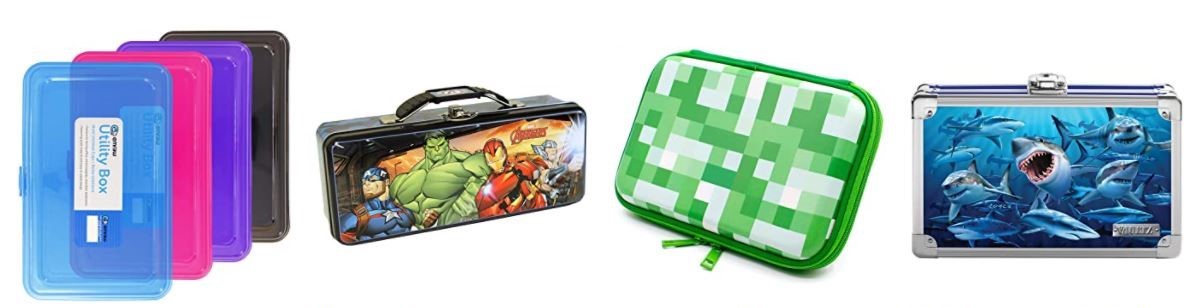}
    }
\caption{Products recommended and results by Amazon for pencil boxes.}
\label{fig:amazon}
\end{figure}

In search for possible gender stereotypes in search result and recommendation, we begin with Amazon.com since it is commonly used by teachers and parents for purchasing classroom materials and search and recommendation are both core tools used to locate products. 
Figure~\ref{fig:Amazon-Kid-result} shows the top result for the query ``pencil boxes for kids''; this product is rather plain and lackluster. The related product recommendations, show in Figure~\ref{fig:Amazon-kid-Rec}, are similar. These results are indeed neutral, not depicting any specific colors or characters, but seem to be missing aspects that would appeal to children.

To further explore Amazon's results, we explore the results for ``pencil box for girls'' and ``pencil box for boys'', two entries in the query suggestions for ``pencil boxes''. 
From Figure \ref{fig:girl} we see that suggestions for girls include pencil boxes that are in shades of pink and shiny and that display representations of unicorns and rainbows. In Figure \ref{fig:boy} we see that instead, pencil boxes for boys portray blue and green colors with superheros, sea creatures, and video game prints. While more appealing to children, we now see the social norms of stereotypical gender aesthetics being displayed in results. This causes an issue as while the gender neutral options are available, seen in Figure \ref{fig:Amazon-kid-Rec}, the gendered recommendations from Figures \ref{fig:girl} and \ref{fig:boy} seem more exciting and likely to be more aesthetically appealing to children, inherently forcing gendered stereotypes on them to get products they like.

\begin{figure}[!ht]
    \subfigure[\label{fig:g_ballet}``ballet" on Google]{
        \includegraphics[width=0.2\linewidth]{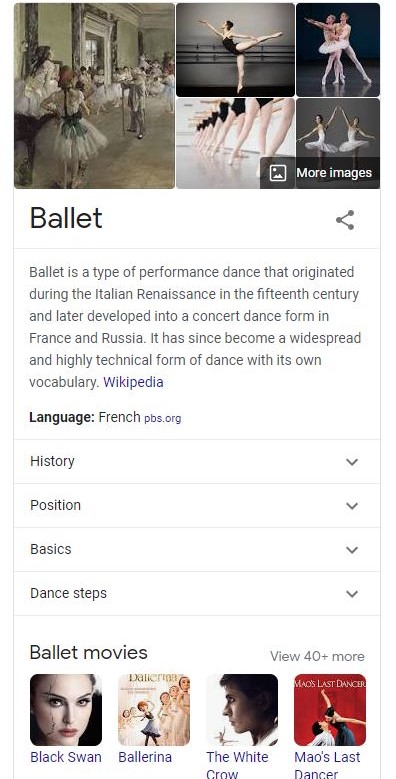}
    }
    \subfigure[\label{fig:g_piret}``pirate" on Google]{
        \includegraphics[width=0.24\linewidth]{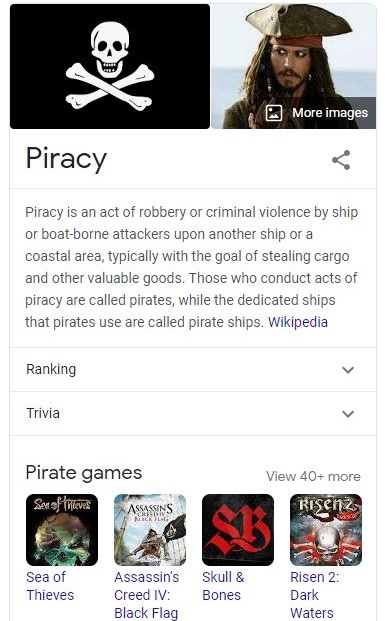}
    }
    \subfigure[\label{fig:k_ballet}``ballet" on Kiddle]{
        \includegraphics[width=0.23\linewidth]{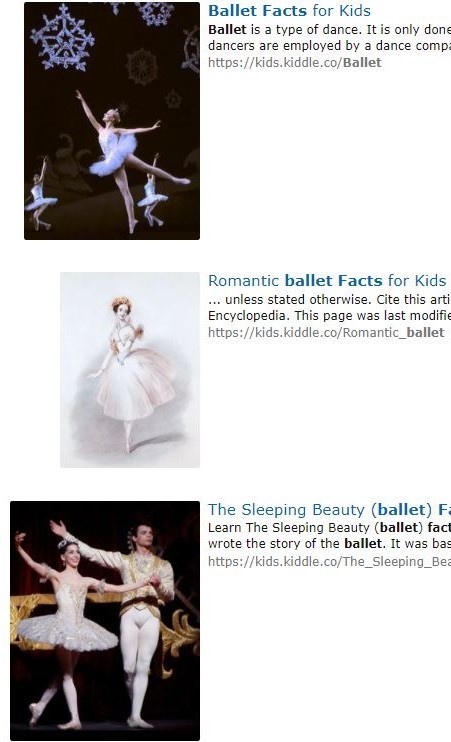}
    }
        \subfigure[\label{fig:k_pire}``pirate" on Kiddle]{
        \includegraphics[width=0.27\linewidth]{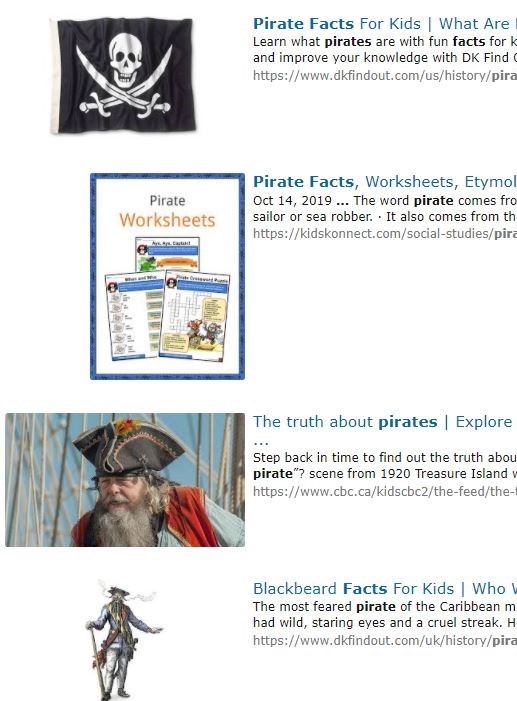}
    }
\caption{Search results for Google and Kiddle.}
\label{fig:search}
\end{figure}

When looking for information, rather than products, search engines are the go to technology for finding information and educational materials for the classroom for both teachers and children. Mainstream search engines like Google aim to serve a general population, while child- and classroom-specific search engines like Kiddle\footnote{https://www.kiddle.co/} are specifically designed to support children in their searches. 
Figure \ref{fig:search} shows results for the queries ``ballet'' and ``pirate'' captured in Figure \ref{fig:search}, as these represent general topic children may search for resources on in the classroom.
In Google's results, we see from Figure \ref{fig:g_ballet} that all but one of the people pictured seem to be female and the only male present is in a seemingly supporting role. The color palette is also very subdued and pastel in nature (prominently pink), connecting to the  societal expectation that the color pink is associated with girls. In contrast, the results in Figure \ref{fig:g_piret} show only male characters, with more intense colors (mainly blacks and reds). Even the suggestions at the bottom of the page differ: ballet suggests movies whereas pirate showcases video games. It is already established that the video game industry is notoriously male-dominated with mostly male-focused storylines \cite{vg_male}.
We see similar trends in specifically child-oriented results displayed by Kiddle: in Figures \ref{fig:k_ballet} and \ref{fig:k_pire}, we see very similar trends regarding the gender of characters and color choices.

These examples are a few of the many we have seen across different platforms that appear to be reinforcing gender stereotypes. Available gender neutral options are nondescript and seemingly promote utility over creativity, which children need to flourish. Thus, pushing children to perpetuate gender stereotypes to express themselves. 
From this preliminary exploration, we posit that gender stereotypes are likely present in search and recommender systems. Considering the power of such systems, this issue is not something that can be overlooked.
We believe information retrieval and education researchers need to take a deeper look into this matter, utilizing the perspective of teachers, to ensure search and recommendation technologies are not contributing to the negative influence of gender stereotypes on children's' development and self-image, particularly in the classroom. 
In the following section, we discussion some possible starting points for research paths to further explore and address gender stereotypes in recommendation and search systems.

\section{Proposed Research Agenda}
Principal 1.2 of the ACM Code of Ethics and Professional Conduct \citep{gotterbarn2018acm}, ``Avoid harm'', requires \emph{proactive} identification and mitigation of potential adverse impacts in computing systems. To address the potential harms arising from gender stereotypes in search and recommendation, we propose research goals that examine this issue through  measurement, source identification, studying impact, and mitigation.

\textbf{Measurement}. To identify the role of search and recommendations in propagating gender stereotypes we must be able to measure their presence. Several metrics have been proposed to measure bias in ranking output \cite{raj2020comparing}, but there are not yet well-established methods for measuring stereotypes. To begin to bridge this gap we propose some research questions that can guide researchers in this area, including: (1) \textit{How do gender stereotypes manifest in result lists in a systematic and detectable way?}; 
(2) \textit{What data can help us to quantify gender stereotypes in systems?}; (3) \textit{What metrics can be used to measure gender stereotypes in search and recommendations?}

\textbf{Cause}. Gender stereotypes are unlikely to originate in a search or recommender system itself, but enter it from society through various channels, including provider input such as item descriptions and metadata, consumer feedback, and algorithm design, among possible other sources \cite{suresh2019framework, ball2021differential}. Bender, Gebru, et al. \citep{Ben:Geb:McM:21b} argued that language models can reinforce and propagate social stereotypes.
To better understand the causes of gender stereotypes in search and recommendations, we propose researchers address the following questions: (1) \textit{To what extent are gender stereotypes present in item descriptions, and what impact does that source have on introducing gender stereotypes in search results and recommendations?}; (2) \textit{To what extent are gender stereotype present in user feedback and what impact does that source have on introducing gender stereotypes in search results and recommendations?}; (3) \textit{Do language models introduce gender stereotypes in search results and recommendations?}

\textbf{Impact}.
The impact of exposing children to gender stereotypes in learning environment through search and recommendations, is still unexplored. While many teachers are aware of gender stereotypes \cite{gray2004perpetuating, lo2015teachers} and practice gender equality in classroom, their experience and perspectives with respect to stereotypes reinforced through search and recommendation systems. Moreover, children can be introduced to gender stereotypes through family \cite{del2019chilean} and their surroundings; we do not yet know to what extent gender stereotypical search results and recommendations exacerbate the impact of early childhood exposure. 
Some seed research questions that may help to address the impact of gender stereotypes in search results and recommendations on learning environments: 
(1) \textit{How does the exposure of gender stereotypes through search and recommendations affect the learning environment in school?}; (2) \textit{Are there identifiable differences in children behavior before and after exposure of stereotypes through search and recommendations?}; (3)\textit{How can we distinguish the impact of early childhood exposure of gender stereotypes introduced by family and other societal forces from learning through search and recommendations?}; (4) \textit{How do search and recommender systems support or impede teacher efforts to provide stereotype-free learning environments?}

\textbf{Mitigation}.
To ensure an equatable learning opportunity for children using search and recommendations, we believe we need to mitigate gender stereotypes that may be reinforced by systems. Some potential questions for reducing or eliminating gender stereotypes in search results and recommendations include: (1)\textit{What design goals can help mitigate gender stereotypes from search results and recommendations?}; (2) \textit{What adjustments can be made to input data to mitigate gender stereotypes from search results and recommendations?}; (3) \textit{What adjustments need to be made to algorithms to build a stereotype-free system for children?}; (4) \textit{How do teachers want search and recommendations to respond while ensuring social equity in the classroom?}; (5) \textit{What role can teachers play in eliminating the impact of stereotypes on children while helping them learn how to interact with the systems?}

\begin{acks}
Work partially supported by NSF Award 1930464 and NSF grant IIS 17-51278.
\end{acks}

\bibliographystyle{ACM-Reference-Format}
\bibliography{sample-base}
\end{document}